\documentclass[11pt,floatfix,amssymb,pra,aps,longbibliography,superscriptaddress,nopacs]{revtex4-2}

\usepackage{comment}
\usepackage[dvipsnames]{xcolor}

\newcommand{\Gopt}{\Gamma_{\mathrm{opt}}}

\newcommand{\Gm}{\Gamma_{\mathrm{m}}}

\newcommand{\Omegam}{\Omega_{\mathrm{m}}}

\newcommand{\Weff}{\Omega_{\mathrm{eff}}}
\newcommand{\ud}{\mathrm{d}}

\renewcommand\Re{\operatorname{Re}}
\renewcommand\Im{\operatorname{Im}}

\newcommand{\La}{\mathrm{L}_1}
\newcommand{\Lb}{\mathrm{L}_\mathrm{c}}
\newcommand{\neff}{\mathrm{n}_{\scriptstyle{\mathrm{eff}}}}
\newcommand{\nth}{\mathrm{n}_{\scriptstyle{\mathrm{th}}}}
\newcommand{\nmin}{\mathrm{n}_{\scriptstyle{\mathrm{min}}}}
\newcommand{\nexc}{\mathrm{n}_{\scriptstyle{\mathrm{exc}}}}
\newcommand{\nBA}{\mathrm{n}_{\scriptstyle{\mathrm{ba}}}}
\newcommand{\Geff}{\Gamma_{\scriptstyle{\mathrm{eff}}}}
\newcommand{\Omegaeff}{\Omega_{\scriptstyle{\mathrm{eff}}}}
\newcommand{\kB}{k_{\scriptscriptstyle{\mathrm{B}}}}
\newcommand{\go}{g_{0}}
\newcommand{\chic}{\chi_\mathrm{c}}
\newcommand{\chim}{\chi_\mathrm{m}}
\newcommand{\chieff}{\chi_\mathrm{eff}}
\newcommand{\aout}{\tilde{a}_{\mathrm{out}}}
\newcommand{\tapout}{\tilde{a}_{\mathrm{out,p}}}
\newcommand{\apout}{a_{\mathrm{out,p}}}
\newcommand{\alphapout}{\alpha_{\mathrm{out,p}}}
\newcommand{\alphapin}{\alpha_{\scriptscriptstyle{0},\mathrm{p}}}
\newcommand{\atot}{\tilde{a}_{\mathrm{tot}}}
\newcommand{\chicp}{\chi_\mathrm{c,p}}

\newcommand{\Sout}{S_{\mathrm{out}}}
\newcommand{\aeff}{a_{\mathrm{eff}}}
\newcommand{\Goptm}{\Gamma_{\mathrm{min}}}
\newcommand{\famp}{A}

\usepackage{amsmath}
\usepackage{graphicx}
\usepackage{epsfig}
\usepackage{tabularx}
\usepackage{color}
\begin{document}

\title{Optical self-cooling of a membrane oscillator in a cavity optomechanical experiment at room temperature}

\author{P. Vezio}
\affiliation{Dipartimento di Fisica e Astronomia, Universit\`a di Firenze, via Sansone 1, I-50019 Sesto Fiorentino (FI), Italy}
\author{M. Bonaldi}
\affiliation{Institute of Materials for Electronics and Magnetism, Nanoscience-Trento-FBK Division,
 38123 Povo, Trento, Italy}
\affiliation{Istituto Nazionale di Fisica Nucleare (INFN), Trento Institute for Fundamental Physics and Application, I-38123 Povo, Trento, Italy}

\author{A. Borrielli}
\affiliation{Institute of Materials for Electronics and Magnetism, Nanoscience-Trento-FBK Division,
 38123 Povo, Trento, Italy}
\affiliation{Istituto Nazionale di Fisica Nucleare (INFN), Trento Institute for Fundamental Physics and Application, I-38123 Povo, Trento, Italy}

\author{F. Marino}
\affiliation{CNR-INO, largo Enrico Fermi 6, I-50125 Firenze, Italy}
\affiliation{INFN, Sezione di Firenze, via Sansone 1, I-50019 Sesto Fiorentino (FI), Italy}

\author{B. Morana}
\affiliation{Institute of Materials for Electronics and Magnetism, Nanoscience-Trento-FBK Division,
 38123 Povo, Trento, Italy}
\affiliation{Dept. of Microelectronics and Computer Engineering /ECTM/DIMES, Delft University of Technology, Feldmanweg 17, 2628 CT  Delft, The Netherlands}

\author{P.M. Sarro}
\affiliation{Dept. of Microelectronics and Computer Engineering /ECTM/DIMES, Delft University of Technology, Feldmanweg 17, 2628 CT  Delft, The Netherlands}

\author{E. Serra}
\affiliation{Istituto Nazionale di Fisica Nucleare (INFN), Trento Institute for Fundamental Physics and Application, I-38123 Povo, Trento, Italy}
\affiliation{Dept. of Microelectronics and Computer Engineering /ECTM/DIMES, Delft University of Technology, Feldmanweg 17, 2628 CT  Delft, The Netherlands}

\author{F. Marin}
\email[Electronic mail: ]{marin@fi.infn.it}
\affiliation{Dipartimento di Fisica e Astronomia, Universit\`a di Firenze, via Sansone 1, I-50019 Sesto Fiorentino (FI), Italy}
\affiliation{European Laboratory for Non-Linear Spectroscopy (LENS), via Carrara 1, I-50019 Sesto Fiorentino (FI), Italy}
\affiliation{INFN, Sezione di Firenze, via Sansone 1, I-50019 Sesto Fiorentino (FI), Italy}
\affiliation{CNR-INO, largo Enrico Fermi 6, I-50125 Firenze, Italy}

\date{\today}

\begin{abstract}
Thermal noise is a major obstacle to observing quantum behavior in macroscopic systems. To mitigate its effect, quantum optomechanical experiments are typically performed in a cryogenic environment. However, this condition represents a considerable complication in the transition from fundamental research to quantum technology applications. It is therefore interesting to explore the possibility of achieving the quantum regime in room temperature experiments. In this work we test the limits of sideband cooling vibration modes of a SiN membrane in a cavity optomechanical experiment. We obtain an effective temperature of a few mK, corresponding to a phononic occupation number of around 100. We show that further cooling is prevented by the excess classical noise of our laser source, and we outline the road toward the achievement of ground state cooling,
\end{abstract}

\maketitle

The research in the field of cavity optomechanics \cite{Aspelmeyer2014,Bowen2015} has gained a lot of momentum in recent years, driven by the observation of quantum phenomena in optically cooled micro- and nanomechanical resonators. This quantum breakthrough is paving the way for integrated systems implementing quantum measurements in sensing devices.
In general, quantum properties of the optomechanical system are hidden or destroyed by thermal noise. As a consequence, most of the quantum optomechanical experiments performed to date have exploited resonators in a cryogenic environment. However, this condition is a major obstacle to making usable sensors. Therefore, a recent branch of research is progressing toward a new generation of optomechanical systems, capable of maintaining quantum behavior, and in particular approaching the mechanical ground state, even at room temperature. A disruptive impact has been achieved by systems based on levitated nanoparticles, whose oscillatory motion in optical tweezers has been cooled down to a phononic occupancy $\bar{n}$ below unity both with a passive scheme exploiting light scattered in a red-detuned cavity mode \cite{Delic2020,Ranfagni2022,Piotrowski2023}, and by measurement-based active feedback \cite{Magrini2021,Tebbenjohanns2021,Kamba2022}. On the opposite side of the mass range, a phononic occupation number of $\bar{n} = 11$ was reported for the pendulum-like motion of the mirrors of the large baseline interferometer LIGO \cite{Whittle2021}. Concerning deformation modes of micro- and nano-devices, active feedback cooling exploiting radiation pressure allowed to cool a vibration mode of a nanometric string down to $\bar{n} = 27$ at room temperature \cite{Guo2019}, and  $\bar{n} = 3.5$ in a liquid nitrogen environment at 77 K \cite{Guo2023}. The same technique led to similar results for a defect mode of a phononic crystal patterned on a SiN membrane, cooled down to $\bar{n} = 20$ \cite{Saarinen2023} in a room temperature experiment. 

In this work we operate with a SiN membrane inside a high finesse optical cavity \cite{Thompson2008} at room temperature, and unlike the cited works we exploit passive, resolved-sidebands cooling. 
This kind of setup is one of the most successful in optomechanics, and allowed to achieve in the past several milestones. It is one of the first systems surpassing the threshold of $\bar{n} <1$ in cryogenic experiments, both by means of sideband cooling \cite{Underwood2015, Peterson2016} and with active feedback \cite{Rossi2018}.     
Having in mind an easier application in future sensing devices, here we chose to realize a robust, self-aligned cavity without fine adjustments and piezoelectric transducers. Our experiment and results are described in the next sections, where we show that we could achieve an effective temperature of a few mK ($\bar{n}$ around 100), a limit well explained by a model including excess laser noise, that is also crucial to provide reliable thermometry.  

\section{Experimental setup}
\begin{figure}[!htb]
   \centering
    \includegraphics[width=0.95\textwidth]{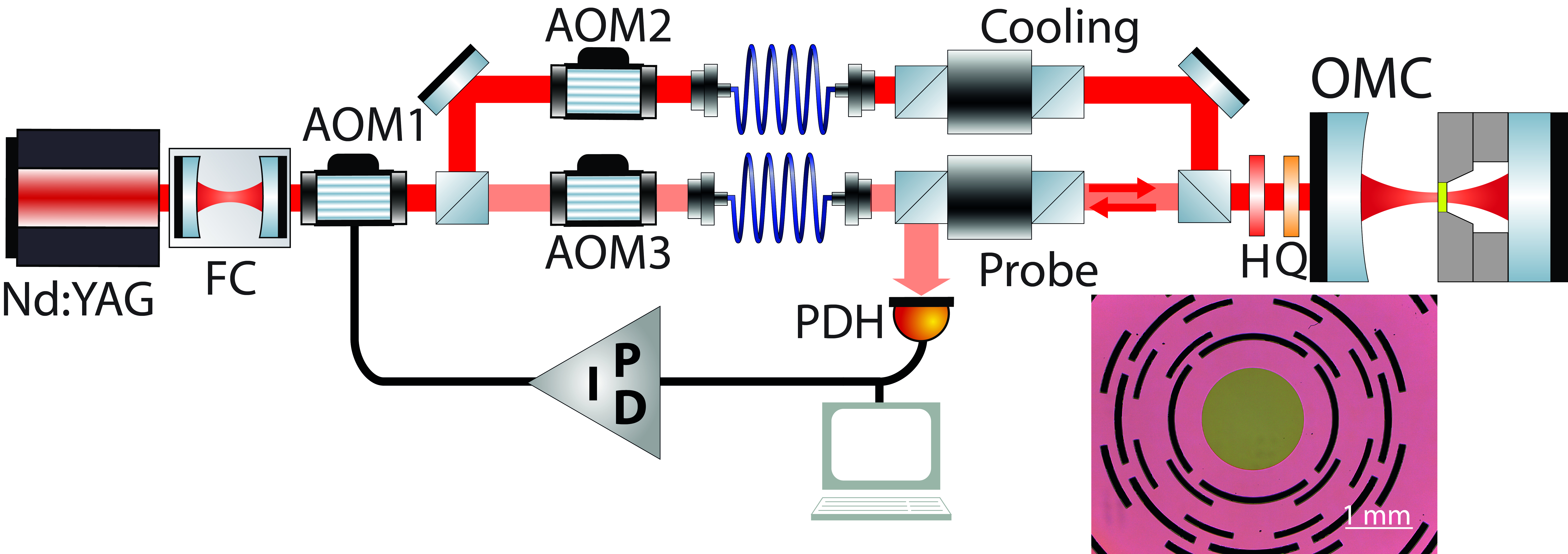}
    \caption{Simplified scheme of the experimental setup. FC: filter cavity. AOM: acousto-optic modulator. OMC: optomechanical cavity. PDH: Pound-Drever-Hall detection. H: half-wave plate. Q: quarter-wave plate.}
    \label{fig_setup}
\end{figure}

A simplified scheme of the experimental setup is sketched in Figure \ref{fig_setup}. The mechanical oscillator is a circular SiN membrane with a thickness of $100\,$nm and a diameter of $1.5\,$mm, equipped with a specific on-chip structure that, working as a ``loss shield'' \cite{Borrielli2014,Borrielli2016,Serra2016,Serra2018}, reduces the coupling between the membrane and the frame and the consequent dissipation losses. In addition, the membrane thickness is reduced at the edge, in order to further decrease the edge losses \cite{Serra2021}. 
In this work we exploit the first two rotationally symmetric drum modes, respectively at $\Omegam/2\pi \simeq 256$ kHz (mode (0,1)), and $\Omegam/2\pi \simeq 593$ kHz (mode (0,2)). The mechanical quality factors $Q$ have been measured from the ring-down of driven oscillations, obtaining respectively $(1.18\pm 0.03) \times 10^7$ and $(0.92 \pm 0.06) \times 10^7$. 

The configuration of the optomechanical cavity is dictated by two requirements: a) it should be self-aligned, without position and tilt adjustments for the membrane and the mirrors, and b) we want to avoid piezoelectric transducers. Due to the latter requirement, by tuning the laser source we have both to optimize the position of the membrane with respect to the cavity standing wave and to find a resonance of the overall cavity. For what concerns the optical alignment, we have to consider that the membrane plane must be orthogonal to the cavity optical axis. To assure it, we have implemented a cavity with a flat end mirror. Silicon spacers between the mirror and the membrane frame, obtained from high quality wafers, guarantee the required parallelism between the surfaces of the mirror and the membrane.  In order to achieve a good optomechanical coupling, we have to vary the position of the membrane with respect to the cavity standing wave, therefore the distance $\La$ between the flat mirror and the membrane should be ideally at least $c/4 \Delta \nu$  ($c$ is the speed of the light, and $\Delta \nu$ is the overall laser tuning range). In our case, $\Delta \nu \simeq 30\,$GHz and we have used two 1 mm thick spacers, giving $\La = 2\,$mm. To assure that a cavity resonance is close to the optimal frequency, we require that the cavity length $\Lb$ is much larger than $\La$. Moreover, it is important that the cavity waist is not too large, to avoid scattering of the light impinging outside the membrane edge. For this purpose, we have chosen a nearly hemifocal configuration, with a concave input coupler having a radius of 50 mm, and a total cavity length of $\Lb = 48\,$mm guaranteed by an invar cylindrical spacer. The cavity finesse, measured with the membrane in working conditions, is around 15400, with an input coupler transmission of 330 ppm.

The cavity is positioned on a cantilever suspension system inside a vacuum chamber, evacuated down to $10^{-6}\,$mbar. The emission of Nd:YAG laser is frequency locked to a 22 cm long filter cavity having a linewidth of 66 kHz. The slow branch of the feedback loop is sent to a piezoelectric transducer (PZT) moving one mirror of the filter cavity, while the fast branch controls the laser frequency by means of its internal PZT. The light transmitted by the filter cavity is tuned by a double-pass acousto-optic modulator (AOM) and split into two beams (the cooling and the probe beams)  whose frequency difference is set by two additional AOMs. The two beams are sent to the experimental bench by optical fibers, overlapped with orthogonal polarizations, and mode-matched to the optomechanical cavity. The reflected probe beam is used for frequency locking the laser to a cavity resonance, by means of a Pound-Drever-Hall (PDH) detection and a servo loop acting on the first AOM, with a locking bandwidth of $\sim 10\,$kHz. For this purpose, the probe beam is phase modulated at 13.3 MHz before coupling it to the fiber.

The same PDH signal is also used to obtain the spectrum of the membrane motion. The power of the probe beam is around $100\, \mu$W, and the signal is converted into frequency fluctuations thanks to calibration tones. The cooling beam power is varied between 0 and $\sim 1\,$mW. 

The frequency spectrum, corrected for the cavity filtering, can be used to deduce the single-photon optomechanical coupling rate $\go$. As described in the next section, in the regime of moderate optomechanical cooling, when the thermal noise is the dominant source of force fluctuations, the peak area of the mechanical resonance can be written as $(\go/2 \pi)^2\,(2\, \neff +1)$ where the phonon occupancy is $\neff = \frac{\kB T}{\hbar \Omegam}\frac{\Gm}{\Geff}$ ($\kB$ is the Boltzmann constant, $T$ the room temperature, $\Gm=\Omegam/Q$ the natural mechanical width, $\Geff$ the measured peak width). Typical values of $\go/2\pi$ are around 2 Hz for both modes.

\section{Model}
The role of excess phase and amplitude noise of the laser fields used for cooling and probing the mechanical oscillators was soon recognized by the optomechanical community, and its contribution to the achievable thermal occupation number was calculated \cite{Rabl2009,Jayich2012,SafaviNaeini2013}. These classical fluctuations are present in the output field both as intrinsic noise, and because the field probes the motion of the mechanical oscillator, in turn driven by the intracavity intensity fluctuations. As a consequence, the output spectrum is modified with respect to the case of coherent fields. The shape of the spectra obtained by heterodyne and by direct intensity detection was derived, as well as their correct use for thermometry \cite{Jayich2012,SafaviNaeini2013,Sudhir2017}. In this section we describe a model of the optomechanical system including classical noise, and calculate the shape of homodyne spectra of the probe field, in order to clearly and correctly interpret our experimental results.        

The linearized evolution equations for the intracavity field operator $\delta\hat{a}$ and the mechanical bosonic operator $\hat{b}$, in the frame rotating with angular frequency $\omega_{\mathrm{L}}=2\pi\nu_{\mathrm{L}}$ ($\nu_{\mathrm{L}}$ is the laser frequency), are \cite{Aspelmeyer2014}
\begin{equation}
\delta \dot{\hat{ a}}=\bigg(  i\Delta-\frac{\kappa}{2}\bigg)\delta\hat{a}+ig_0\,\alpha(\hat{b}+\hat{b}^\dag)+\sqrt{\kappa}\,\delta\hat{a}_{\mathrm{in}}
\label{dta}
\end{equation}
\begin{equation}
\dot{\hat{b}}=\left(-i\Omegam-\frac{\Gm}{2}\right)\hat{b}+ig_0(\alpha^* \delta\hat{a}+\alpha \delta \hat{a}^{\dag})+\sqrt{\Gm}\,\hat{b}_{\mathrm{in}}
\label{dtb}
\end{equation}
where $\Delta=\omega_{\mathrm{L}}-\omega_{\mathrm{c}}$ is the detuning with respect to the cavity resonance frequency $\omega_{\mathrm{c}}$, $\kappa$ and $\Gm$ are the optical and mechanical decay rates, 
$\alpha$ is the intracavity mean field and the input terms are defined below.

In the Fourier space, equations (\ref{dta},\ref{dtb}) can be written as
\begin{equation}
\frac{1}{\chic} \tilde{a}=  ig_0 \,\alpha \left(\tilde{b}+\tilde{b}^{\dag}\right)
+\sqrt{\kappa}\; \tilde{a}_{\mathrm{in}} 
\label{Fdta}
\end{equation}
\begin{equation}
\frac{1}{\chim}  \tilde{b}=  ig_0 \left(\alpha^* \tilde{a}+\alpha \tilde{a}^{\dag}\right)
+\sqrt{\Gm}\; \tilde{b}_{\mathrm{in}} 
\label{Fdtb}
\end{equation}
where we use $\tilde{O}$ to indicate the Fourier transformed of the operator $\hat{O}$, and  $\tilde{O}^{\dag}$ for the Fourier transformed of $\hat{O}^{\dag}$, such that $\left(\tilde{O}(\omega)\right)^{\dag}=\tilde{O}^{\dag}(-\omega)$. The optical and mechanical susceptibilities are defined as 
\begin{gather}
\chim\left(\omega\right)=\frac{1}{-i
\left(\omega-\Omegam\right)+\Gm/2}\\
\chi_\mathrm{c}\left(\omega\right)=\frac{1}{-i
\left(\omega+\Delta\right)+\kappa/2}
\end{gather}
and the intracavity mean field $\alpha$ is related to the input field $\alpha_{0}$ by the equation $\alpha = \sqrt{\kappa} \chic(0) \alpha_{0}$.

Replacing Eq. (\ref{Fdta}) in the (\ref{Fdtb}) and neglecting the counter-rotating terms, the solution for $\tilde{b}$ can be written as
\begin{equation}
\frac{1}{\chieff}  \tilde{b}=  ig_0 \sqrt{\kappa} \left(\alpha^* \chic(\omega) \tilde{a}_{\mathrm{in}} +\alpha \chic^*(-\omega) \tilde{a}^{\dag}_{\mathrm{in}} \right)
+\sqrt{\Gm}\; \tilde{b}_{\mathrm{in}}   
\label{eq_btilde1}
\end{equation}
where the effective susceptibility $\chieff$ is defined as 
\begin{equation}
\frac{1}{\chieff} = \frac{1}{\chim}+g_0^2 \,|\alpha|^2 \big(\chic(\omega) - \chic^* (-\omega)\big) \simeq i\left(\Omegaeff-\omega\right)+\frac{\Geff}{2}  \,.
\end{equation}
The resonance frequency is shifted to $\Omegaeff = \Omegam + g_0^2 |\alpha|^2 \Im \big[\chic(\Omegam) - \chic^* (-\Omegam)\big]$ by the optical spring effect, and red detuned radiation provides optical damping with rate 
\begin{equation}
\Gopt = 2 g_0^2 \, |\alpha|^2 \Re \big[\chic(\Omegam) - \chic^* (-\Omegam)\big]
\label{eq_Gopt}
\end{equation}
yielding a total width $\Geff = \Gm + \Gopt$.

The input field fluctuations including the classical extra phase and amplitude noise can be written as 
\begin{equation}
\delta\hat{a}_{\mathrm{in}} = \delta\hat{a}_{\mathrm{v}} + \alpha_0\, \left( i \phi + \epsilon \right)
\label{eq_ain}
\end{equation}
where the phase and relative amplitude fluctuations are given by the classical, real stochastic variables $\phi$ and $\epsilon$, which we assume to be uncorrelated. The 
input noise operators are characterized by the correlation functions
\begin{eqnarray}
\langle\hat{a}_{\mathrm{v}}(t)\hat{a}_{\mathrm{v}}^{\dag}(t')\rangle & = & \delta(t-t')
\label{noise1} \\
\langle\hat{a}_{\mathrm{v}}^{\dag}(t)\hat{a}_{\mathrm{v}}(t')\rangle & = & 0
\label{noise2}  \\
\langle\hat{b}_{\mathrm{in}}(t)\hat{b}_{\mathrm{in}}^{\dag}(t')\rangle & = & (\nth+1)\,\delta(t-t')
\label{noise3}    \\
\langle\hat{b}_{\mathrm{in}}^{\dag}(t)\hat{b}_{\mathrm{in}}(t')\rangle & = & \nth\,\delta(t-t')
\label{noise4}
\end{eqnarray}
where $\nth$ is the thermal occupation number of the thermal bath, that in our case is at room temperature. 

Replacing the expressions of the input field noise in Eq. (\ref{eq_btilde1}), we derive 
\begin{equation}
\begin{split}
\frac{1}{\chieff}  \tilde{b}=  ig_0 \sqrt{\kappa} \left(\alpha^* \chic(\omega) \tilde{a}_{\mathrm{v}} +\alpha \chic^*(-\omega) \tilde{a}^{\dag}_{\mathrm{v}} \right) 
- g_0 \sqrt{\kappa} \alpha_0 \Big(\alpha^* \chic(\omega)  -\alpha \chic^*(-\omega) \Big) \tilde{\phi}\\
+i g_0\sqrt{\kappa} \alpha_0 \Big(\alpha^* \chic(\omega)  +\alpha \chic^*(-\omega) \Big) \tilde{\epsilon} + \sqrt{\Gm}\; \tilde{b}_{\mathrm{in}}  \, .
\label{eq_btilde2}
\end{split}
\end{equation}
The effective thermal occupation number can be calculated as $\neff = \int\int \langle  \tilde{b}^{\dag}(\omega')\tilde{b}(\omega) \rangle \frac{\ud \omega'}{2\pi}\frac{\ud \omega}{2\pi}$, obtaining in the weak coupling limit
\begin{equation}
\neff = \,\frac{\Gm}{\Geff}\nth + \frac{\Gopt}{\Geff} \left( \nBA + \nexc \right)
\label{eq_neff}
\end{equation}
where the contribution due to the quantum backaction is \cite{Aspelmeyer2014}
\begin{equation}
\nBA = \left(\frac{(\kappa/2)^2+(\Delta-\Omegam)^2}{(\kappa/2)^2+(\Delta+\Omegam)^2}-1\right)^{-1}
\end{equation}
and the contributions of the excess phase and amplitude noise sources, proportional to their spectral densities $S_{\phi \phi}$ and $S_{\epsilon \epsilon}$, is 
\begin{equation}
\nexc = \frac{\kappa^2 g_0^2 \alpha_0^4}{\Gopt}   \Big( \big|\chic^*(0) \chic (\Omegam)-\chic(0) \chic^* (-\Omegam)\big|^2 \, S_{\phi\phi} +  \big|\chic^*(0) \chic (\Omegam)+\chic(0) \chic^* (-\Omegam)\big|^2 \, S_{\epsilon \epsilon} \Big)   \, .
\end{equation}

\begin{figure}[!htb]
    \centering
    \includegraphics[width=0.8\textwidth]{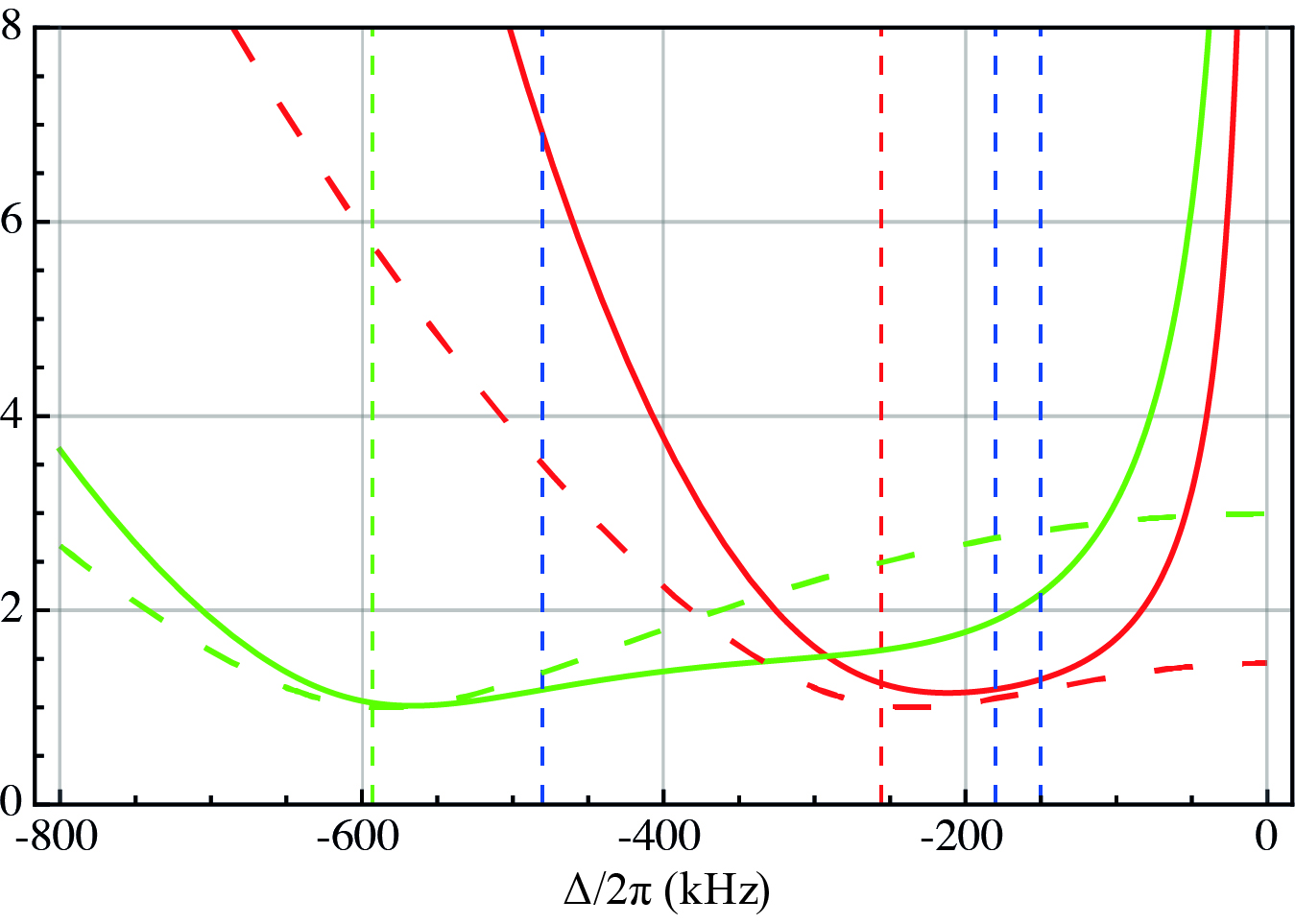}
    \caption{Dimensionless functions $1/\cos \theta$ (dashed lines) and $\famp$ (solid lines), described in the text, plotted as a function of the detuning $\Delta/2\pi$ for $\kappa/2\pi = 204\,$kHz, and $\Omegam/2\pi$ = 256 kHz (red curves) and 593 kHz (green curves). Vertical dotted lines are plotted in correspondence of $\Omegam/2\pi$ (green and red lines) and of the different values of detuning used in the present work (blue lines at -480 kHz, -180 kHz, and -150 kHz).  }
    \label{fig_f}
\end{figure}

Using Eq. (\ref{eq_Gopt}) to replace $\alpha_0$ in the above expression, the excess occupation number can be written in a simple form, useful for the comparison with the experimental results, as
\begin{equation}
\nexc =\Gopt\,\frac{ \Omegam^2}{4 g_0^2}\,\Big( \frac{1}{\cos^2 \theta}\,S_{\phi\phi}\,+\famp^2 S_{\epsilon\epsilon} \Big)
\label{eq_nexc}
\end{equation}
where 
$\theta = \arg \left[ \chic (\Omegam)- \chic^* (-\Omegam) \right]$ and  
\begin{equation}
\famp = \frac{|\chic^*(0) \chic (\Omegam)+\chic(0) \chic^* (-\Omegam)|}{\Omegam |\chic (0) |^2 \Re \left[\chic (\Omegam) - \chic^* (-\Omegam) \right]}  \, .
\end{equation}
The functions $1/\cos \theta$ and $\famp$ are
plotted in Fig. \ref{fig_f} for our experimental parameters $\kappa$ and $\Omegam$. We note that in the strongly resolved sideband regime ($\kappa \ll \Omegam$) and for $\Delta = -\Omegam$, we have $\cos \theta \simeq 1$ and $\famp \simeq 1$. 

In $\neff$, the contribution of the thermal noise (first term on the right-hand side in Eq. (\ref{eq_neff})) is proportional to the inverse of the input power (since $\Geff \simeq \Gopt \propto \alpha_0^2$), while the contribution of the excess noise (rightmost term in Eq. (\ref{eq_neff})) increases linearly with the input power. We can emphasize it by writing $\neff$ as 
\begin{equation}
\neff = 0.5 \,\nmin \left( \frac{\Gopt}{\Goptm}+  \frac{\Goptm}{\Gopt}\right) + \nBA  
\end{equation}
where
\begin{equation}
\nmin = \frac{\Omegam \sqrt{\Gm \nth }}{g_0} \sqrt{\frac{S_{\phi\phi}}{\cos^2 \theta}+\famp^2 S_{\epsilon\epsilon}}  \, .
\label{eq_nmin}
\end{equation}
Varying the input power, the minimum occupancy is achieved when thermal noise and phase noise equally contribute with $0.5 \,\nmin$, in correspondence with the optimal optical width
\begin{equation}
\Goptm =\frac{2 g_0 \sqrt{\Gm \nth}}{\Omegam} \left(\frac{S_{\phi\phi}}{\cos^2 \theta}+\famp^2 S_{\epsilon\epsilon}\right)^{-\frac{1}{2}}   \, .
\end{equation}

The state of the oscillator is measured through the output field, which is given by the input-output relation $a_{\mathrm{out}} = -\sqrt{\kappa} a +a_{\mathrm{in}} $. For the field fluctuations, replacing Eq. (\ref{Fdta}) in this latter relation, we can write
\begin{equation}
\aout = -i g_0 \,\alpha \chic(\omega) \sqrt{\kappa} \left( \tilde{b} + \tilde{b}^{\dag} \right) + \big(1-\kappa \chic (\omega) \big) \tilde{a}_{\mathrm{in}}  
\label{eq_aout}
\end{equation}
and for the mean field $\alpha_{\mathrm{out}} = \left(1-\kappa \chic(0)\right) \alpha_0$.

In order to gather information on the displacement of the oscillator, we analyze a quadrature of the probe field. In particular, in the case of PDH detection, the phase modulation sidebands act as a local oscillator to extract the phase quadrature of the resonant carrier field of the probe beam.  
We assume that all the fields (cooling field, probe, and local oscillator) are derived from the same laser source and are affected by the same excess phase and relative amplitude noise.

For a general quadrature detection, the total detected fluctuations are proportional to
\begin{equation}
    X_{\mathrm{out}}= \frac{1}{2} \left( e^{-i (\theta_{\mathrm{LO}} + \phi)}\,\apout \, + h.c. \right) \, \simeq \, \frac{1}{2} \left(e^{-i \theta_{\mathrm{LO}}}  \atot  + h.c. \right)
\end{equation}
where $\theta_{\mathrm{LO}}$ is the phase of the local oscillator, $\atot = \tapout-i \alphapout \phi\,$ and we use the subscript p for the probe field. Neglecting the quantum noise of the probe and using Eqs. (\ref{eq_ain}) and (\ref{eq_aout}), the total detected field, after some algebra and noting in particular that $\chic^{-1}(0)-\chic^{-1}(\omega) = i\omega$, can be written as 
\begin{equation}
\atot = -i \alphapin \kappa \chicp(\omega) \chicp (0) \left(i \omega \tilde{\phi} + g_0 (\tilde{b}+\tilde{b}^{\dag})\right) + \alphapin \Big(2-\kappa \chicp (0) -\kappa \chicp(\omega) \Big)\tilde{\epsilon}
\end{equation}
and the detected field quadrature as 
\begin{equation}
        X_{\mathrm{out}}= -\frac{4}{\kappa}\alphapin \, C(\omega) \left(i \omega \tilde{\phi} + g_0 (\tilde{b}+\tilde{b}^{\dag})\right) + D(\omega) \tilde{\epsilon}
\end{equation}
where 
\begin{equation}
C(\omega) = i \frac{\kappa^2}{8} \left(\chicp(\omega) \chicp (0) e^{-i \theta_{\mathrm{LO}}} - \chicp^*(-\omega) \chicp^* (0) e^{i \theta_{\mathrm{LO}}} \right)  
\label{eq_C}
\end{equation}
and
\begin{equation}
D(\omega) =\left(1-\frac{\kappa}{2}\chicp(\omega) -\frac{\kappa}{2}\chicp (0)\right) e^{-i \theta_{\mathrm{LO}}} +\left(1-\frac{\kappa}{2} \chicp^*(-\omega)-\frac{\kappa}{2} \chicp^* (0) \right)e^{i \theta_{\mathrm{LO}}}    \, . 
\end{equation}

In the particular case of a phase quadrature detection with a resonant probe field (i.e., for  $\Delta_{\mathrm{p}} = 0$ and $\theta_{\mathrm{LO}}=\pi/2$), as, e.g., in the signal of a PDH detection, Eq. (\ref{eq_C}) simplifies to 
$C(\omega) = \frac{ \kappa/2}{\kappa/2-i\omega}$,
giving to the usual cavity filtering function, and $D(\omega) = 0$. 
In the following, we consider this experimental situation, and in particular, since  $D(\omega) = 0$, we can neglect the correlation between the mechanical fluctuations and the amplitude noise of the probe and the local oscillator. 

It is useful to separate in $(\tilde{b}+\tilde{b}^{\dag})$ the term proportional to $\tilde{\phi}$ from the expressions in the absence of phase noise. 
Using also Eq. (\ref{eq_btilde2}), we thus write
\begin{equation}
\begin{split}
X_{\mathrm{out}} = -\frac{4}{\kappa}\alphapin \, C(\omega)\Bigg[\Bigg(i \omega + g_0^2 \alpha_0^2 \kappa \,\Big(\chic^*(0) \chic (\omega)-\chic(0) \chic^* (-\omega)\Big) \Big(-\chieff(\omega)+\chieff^*(-\omega)\Big)\Bigg) \tilde{\phi}
\\ +g_0 (\tilde{b}+\tilde{b}^{\dag})_{\mathrm{no\, \phi}} \Bigg] \, .
\end{split}
\label{eq_Xout}
\end{equation}
The symmetrized spectrum of $ X_{\mathrm{out}}$, defined as $\Sout = \frac{1}{2}\int \frac{\ud \omega'}{2\pi}\langle X_{\mathrm{out}}^{\dag}(\omega')X_{\mathrm{out}}(\omega)+X_{\mathrm{out}}^{\dag}(\omega')X_{\mathrm{out}}(-\omega) \rangle$, can be calculated from Eq. (\ref{eq_Xout}). A simple and clear form is obtained in the weak coupling limit (implying $\Geff \ll \kappa, \Omegam$), by replacing $\omega \to \Omegam$ inside the square brackets, everywhere except in $\chieff$, and neglecting the terms proportional to $\chieff(\omega) \chieff (-\omega)$. After some algebra, the output spectrum turns out to be proportional to
\begin{equation}
\Sout\,\propto\,c + |C(\omega)|^2\,\left[2 g_0^2  \left( \neff+\frac{1}{2} \right) \mathcal{L} - 2  g_0^2 \,\nexc  2 \cos \theta \Big(\cos \theta\,\mathcal{L}-  \sin \theta \,\mathcal{D} \Big)\right]
\label{eq_Sout}
\end{equation}
where $c$ is a constant accounting for the vacuum noise not entering the cavity (due, e.g., to the limited efficiency), 
and the Lorentzian and dispersive shapes $\mathcal{L}$ and $\mathcal{D}$ are respectively
\begin{eqnarray}
    \mathcal{L} = \frac{\Geff}{2}\left( |\chieff (\omega)|^2 + |\chieff (-\omega)|^2 \right) \\
    \mathcal{D} =  (\omega-\Weff)|\chieff (\omega)|^2 +(-\omega-\Weff) |\chieff (-\omega)|^2 
\end{eqnarray}
An additional calibration tone, i.e., a coherent phase modulation at a frequency far from $\Omegam$, yields an additional peak in $\Sout$ which is used to determine the overall scale constant in the experimental spectrum and thus obtain it in the form given written at the right-hand side of Eq. (\ref{eq_Sout}).

\section{Experimental results}

\begin{figure}[!htb]
    \centering
    \includegraphics[width=0.8\textwidth]{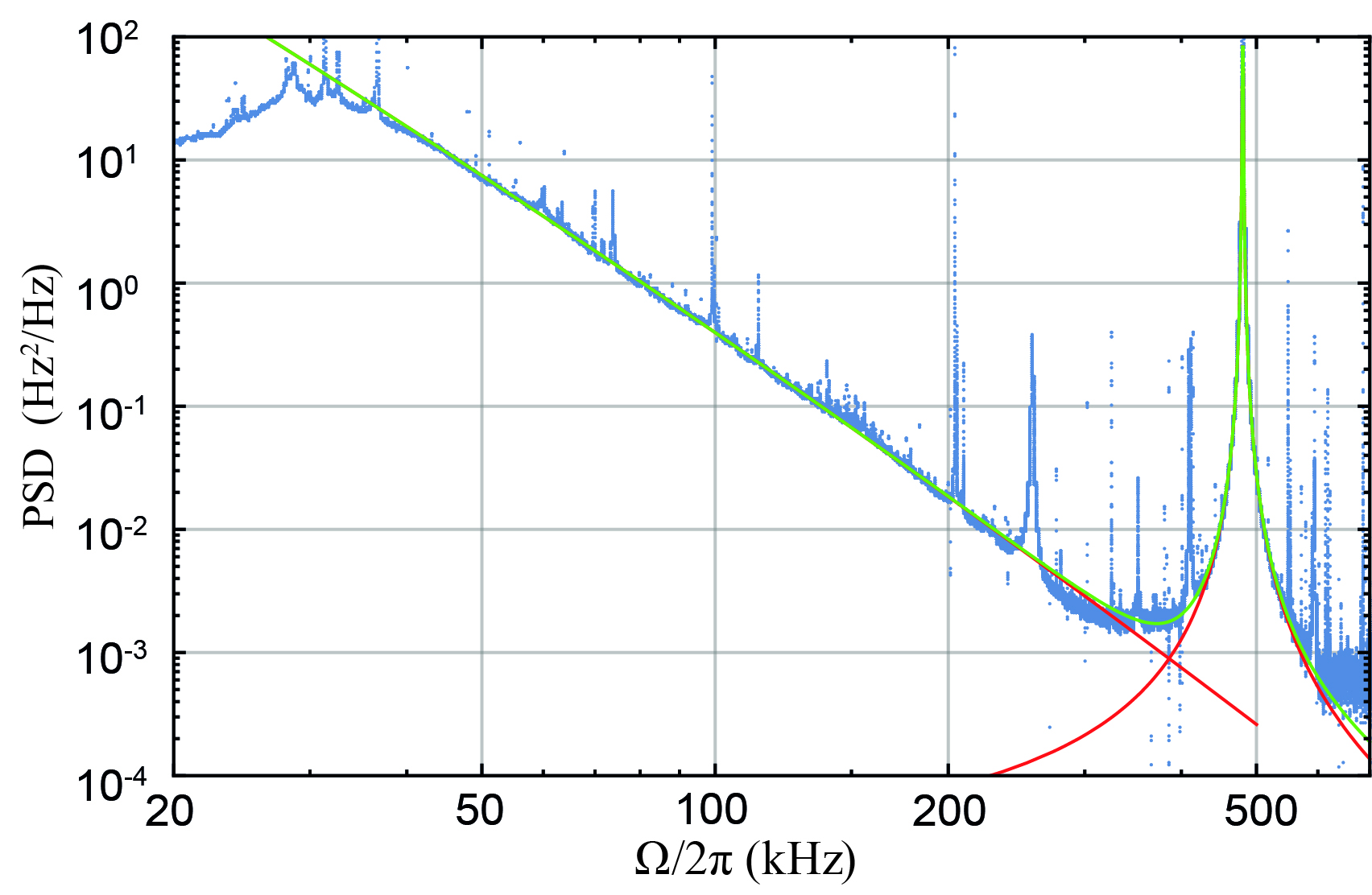}
    \caption{Spectrum (Power Spectral Density) of the PDH signal, calibrated in terms of frequency fluctuations. Red curves: fits of the low-frequency background and of the beat note between probe and cooling beam. Green curve: overall background given by the sum of the two fitted curves.}
    \label{fig_spettro_largo}
\end{figure}
Observing the detected spectrum in a wide frequency range (Fig. \ref{fig_spettro_largo}), we can notice two relevant features: the tail of low-frequency fluctuations, and the beat note between cooling and probe fields. The former is due both to frequency noise and to low-frequency mechanical modes of the oscillator device, belonging to the frame and to the internal filtering structure. We have observed that our filter cavity is efficient in reducing the frequency noise above $\sim 100\,$kHz, but at a lower frequency, on the contrary, the noise is increased. The noise tail gives a contribution to the background beneath the peak of the first membrane mode, at $\sim 260\,$kHz. The beat note is due to residual percolation between the two fields on the reflected path, in spite of their orthogonal polarizations. This beat note is minimized using the half- and quarter-wave plates before the cavity to compensate for its birefringence, however, the residual remains relevant. The two spectral structures are fitted with phenomenological shapes on frequency intervals outside the regions of the membrane modes, and the fitting functions are subtracted from the experimental spectra in order to minimize the residual background.

\begin{figure}[!htb]
    \centering
    \includegraphics[width=0.95\textwidth]{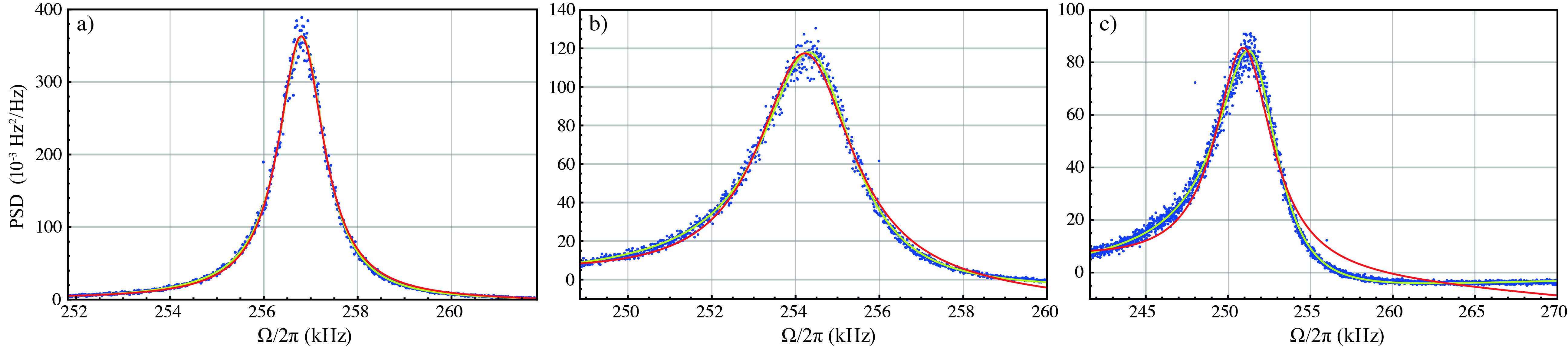}
    \caption{ Spectra of the resonance peak corresponding to the first membrane mode, acquired for a detuning of the cooling beam of $\Delta/2\pi = -480\,$kHz. Panels a), b) and c) correspond to increasing cooling power, giving widths of respectively $\Geff/2\pi = 1.2\,$kHz, $\Geff/2\pi = 2.7\,$kHz and $\Geff/2\pi = 9\,$kHz. Dots: experimental data. Green solid line: fit with the sum of a Lorentzian and a dispersive shape (Eq. (\ref{eq_fit})). Red solid line: fit with a Lorentzian shape (Eq. (\ref{eq_fit}) with $a_3 = 0$). }
    \label{fig_lor1}
\end{figure}
We now focus on the first membrane mode, appearing in the spectra of the PDH detection. At low cooling power, the resonance peak is well fitted by a Lorentzian shape (Fig. \ref{fig_lor1}a), which broadens at increasing cooling power. This broadening is accompanied at the beginning by a reduction of the peak area, but at high power the frequency noise plays a relevant role. The mode heats up again and, if $\cos \theta$ is not too small (depending on the detuning of the cooling field), the lineshape is well fitted by the sum of a Lorentzian and a dispersive curves as predicted by Eq. (\ref{eq_Sout}). A clear example is shown in Fig. \ref{fig_lor1}c, for $\Delta = -480\,$kHz. 

The calibrated output spectra, such as those shown in Fig. \ref{fig_lor1}, are fitted with the function 
\begin{equation}
a_0 + a_1\,\omega + |C(\omega)|^2(a_2  \mathcal{L} + a_3  \mathcal{D})
\label{eq_fit}
\end{equation}
where $a_i$ are free constant parameters and $\kappa$ is fixed at the value measured independently. The term $(a_1\,\omega)$ accounts for the residual background, remaining after the described subtraction of the low-frequency and beat note structures.
We then calculate $\aeff = a_2 + a_3/ \tan \theta $, which according to Eq. (\ref{eq_Sout}) is equal to $\aeff = g_0^2  (2\,\neff+1)$. $\theta$ is calculated from independently measured system parameters. Finally, we fit $\aeff$ vs $\Geff$ with the function $\aeff = b_1/\Geff+b_2 \Geff$, with $b_1$ and $b_2$ as free parameters, a behaviour that should hold for $\Gopt \simeq \Geff$ and $\nmin \gg 1, \nBA$.   
At low cooling power, the thermal noise dominates and according to Eq. (\ref{eq_neff}) we have $b_1 = 2 g_0^2 \,\Gm \nth$. Since $\Gm$ and $\nth$ are determined independently (the latter, assuming that the oscillator is at room temperature), we can evaluate the optomechanical coupling rate $g_0$. The optimal width is derived as $\Goptm = \sqrt{b_1/b_2}$ and the minimum occupation number as $\nmin = 2 \Gm \nth \sqrt{b_2/b_1}$.

\begin{figure}[!htb]
    \centering
    \includegraphics[width=1\textwidth]{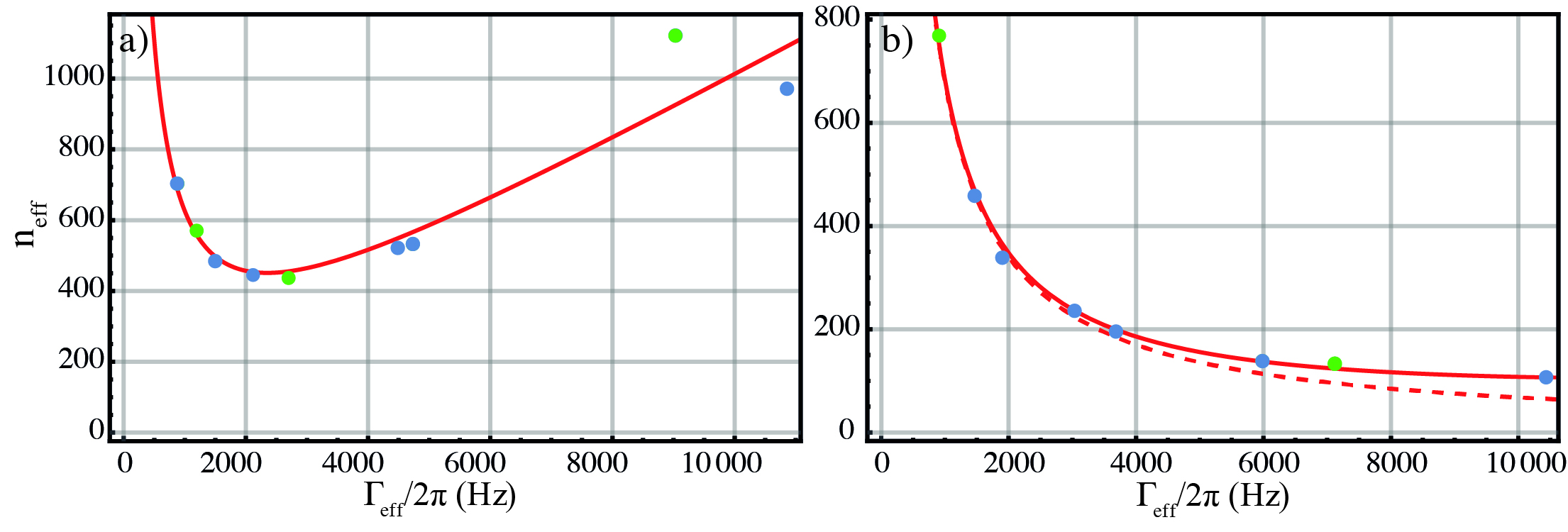}
    \caption{Dots: area of the resonance peak of the (0,1) mode (panel a) and (0,2) mode (panel b) of the membrane, calibrated in terms of thermal occupancy, at increasing cooling power, plotted as a function of the peak width. Green dots correspond to the spectra shown in Figs. \ref{fig_lor1} and \ref{fig_modo02a}. Red solid line: fit with the function $b_1/\Geff+b_2 \Geff$. Dashed line: $b_1/\Geff$ contribution to the fit. }
    \label{fig_neff1}
\end{figure}
In Fig. \ref{fig_neff1} we plot $\aeff/2 g_0^2 \equiv \neff$, for a detuning of the cooling beam of -480 kHz. With the described procedure, we derive $g_0/2\pi = 2.1 \pm 0.1\,$Hz and $\nmin = 450 \pm 35$, at the optimal effective width of $\Goptm/2\pi = 2.35\,$kHz. The acquired spectrum corresponding to the lowest occupancy is shown in Fig. \ref{fig_lor1}b.

The excess occupation number $\nexc$ is produced by both phase and amplitude excess noise. The coefficient of the dispersive shape, which is just sensitive to phase noise, allows distinguishing the two sources. According to Eq. (\ref{eq_Sout}), the contribution of the phase noise to $b_2 \Geff$ should be equal to $a_3/\sin 2\theta$, while a larger $b_2$ can be attributed to the amplitude noise. In our case, we find $b_2 \Geff/a_3 = -2.3 \pm 0.4$, in good agreement with the calculated $1/\sin 2\theta  = -1.84$. As a consequence, we can ascribe $\nexc$ to the excess phase noise.
Using Eq. (\ref{eq_nmin}), we deduce a frequency noise of $S_{\nu\nu} = (\omega/2\pi)^2 S_{\phi\phi} = (2.2 \pm 0.4) \times 10^{-2}\,\mathrm{Hz}^2/\mathrm{Hz}$, around the mode eigenfrequency of $\sim 250\,$kHz. For the relative amplitude noise, we infer an upper limit of $S_{\epsilon\epsilon} < 2 \times 10^{-14}\,\mathrm{Hz}^{-1}$. 

Following the behaviour of  
$1/\cos \theta$ shown in Fig. \ref{fig_f}, we infer that the minimum occupation number is potentially reduced by a factor of 4.4, i.e., down to $\sim 100$, when the detuning is close to $\Delta \simeq -\Omegam \simeq 2\pi\times 260\,$kHz. This working point implies two technical issues in the measurement of the phonon number. Firstly, the beat note between the cooling and probe fields is close to the mechanical peak, hindering its accurate analysis. Secondly, here $|\sin \theta|$ is small, therefore it is difficult to determine the weight of the dispersive contribution in the spectral shape and consequently the correction to be applied to the Lorentzian amplitude. We have however acquired sets of spectra at varying cooling power, for detunings of $\Delta/2\pi = -150\,$kHz and  $\Delta/2\pi = -180\,$kHz. The deduced optomechanical coupling rate is now $g_0/2\pi = 2.6\,$Hz, and the slight increase with respect to the previous value can be attributed to a different position of the membrane with respect to the cavity standing wave, caused by thermal drifts (these datasets were acquired on a different day than the previous ones). The inferred frequency noise is $S_{\nu\nu}  = (2.2 \pm 1.3) \times 10^{-2}\,\mathrm{Hz}^2/\mathrm{Hz}$ for the data at $\Delta/2\pi = -150\,$kHz, and  $S_{\nu\nu}  = (3.0 \pm 1.4) \times 10^{-2}\,\mathrm{Hz}^2/\mathrm{Hz}$ for the data at $\Delta/2\pi = -180\,$kHz, while the minimum occupation number is respectively $\nmin = 120 \pm 70$ and $\nmin = 130 \pm 60$. Even with the anticipated low accuracy, these results agree with the previously reported data, confirming the overall self-consistency of our modeling.
      
\begin{figure}[!htb]
    \centering
    \includegraphics[width=1\textwidth]{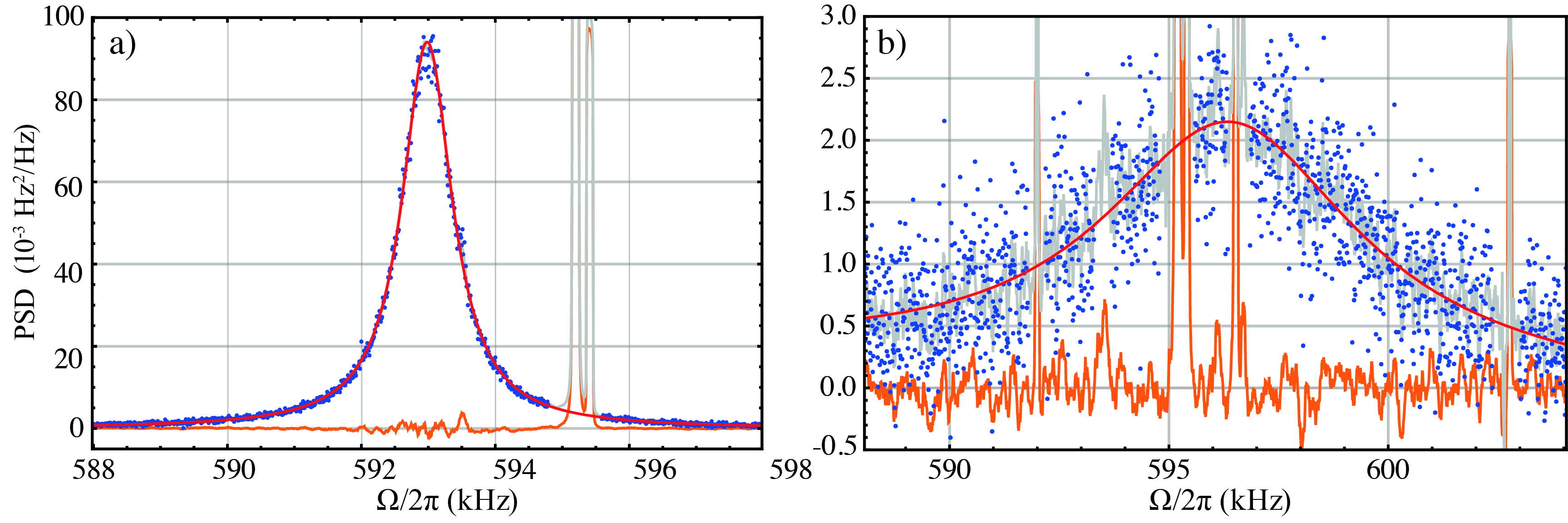}
    \caption{Spectra of the resonance peak corresponding to the (0,2) membrane mode, acquired for a detuning of the cooling beam of $\Delta/2\pi = -480\,$kHz. Panels a), and b) correspond to increasing cooling power, giving widths of respectively $\Geff/2\pi = 900\,$Hz and $\Geff/2\pi = 7.1\,$kHz. Blue dots: experimental data used for the fit. Light gray line: 10 points (equivalent to 100 Hz) moving average of the complete set of experimental data. Few spurious peaks are removed from the set used for the fit. Red solid line: fit with a Lorentzian shape. Orange dots: residuals of the fit (10 points moving average).}
    \label{fig_modo02a}
\end{figure}
We now consider the (0,2) mode of the membrane. Thanks to its higher resonance frequency (about 593 kHz), it has a lower back-action limited occupation number $\nBA$ and it is less sensitive to low-frequency technical noise. However, cavity filtering reduces the signal-to-noise ratio in the PDH detection. An example of the resonance peak acquired at moderate cooling is shown in Fig. \ref{fig_modo02a}a, for a detuning of $\Delta/2\pi = -480\,$kHz. The dispersive contribution to the lineshape cannot be singled out. Even at the highest cooling power that still allows the resonance peak to be reliably extracted from the wings of the beat note, the purely Lorentzian shape gives the best fit to the experimental data, as shown in Fig. \ref{fig_modo02a}b and confirmed by the lack of visible systematic behavior in the fit residuals. The peak area, calibrated as before in terms of occupation number, is plotted in Fig. \ref{fig_neff1}b as a function of the peak width $\Geff$, and fitted with the function $b_1/\Geff+b_2 \Geff$. We infer $g_0/2\pi = 1.74 \pm 0.02\,$Hz and $\nmin = 104 \pm 14$, at the optimal effective width of $\Goptm/2\pi = 13 \pm 2\,$kHz. 

The lack of a dispersive component in the peak lineshape indicates that the frequency noise plays a negligible role, and $\nexc$ should be imputed to the excess amplitude noise. Using Eq. (\ref{eq_nmin}) we infer $S_{\epsilon\epsilon}  = (1.6 \pm 0.2) \times 10^{-14}\,\mathrm{Hz}^{-1}$. According to the behaviour of $\famp$, an optimized detuning ($\Delta \simeq -\Omegam$) would reduce $\neff$ by $20 \%$.

\section{Conclusions}
We have optically cooled two low-frequency vibration modes of a membrane at room temperature, exploiting self cooling in a cavity optomechanical setup with a red-detuned cooling beam in the resolved sideband regime. We have chosen to build a robust monolithic cavity, self-aligned, without the necessity of mechanical adjustments and piezoelectric transducers. For both modes we have achieved a phononic occupation number around 100, corresponding to effective temperatures of respectively 1.5 mK and 3 mK for the (0,1) and (0,2) modes, and effective quality factors between 30 and 50. We have shown that this performance is limited by the laser excess noise. Without changing other parameters, lower occupation numbers could be obtained with oscillators exhibiting higher quality factor, but achieving an occupancy around 1 would require $Q \sim 10^{11}$, well above the state of the art for room temperature membrane oscillators. 

The excess amplitude noise can be reduced using an additional noise eater. In \cite{Pontin2018} we report an active stabilization that reduces the intensity fluctuations at a level that is 3 dB above shot noise for a laser power of 24 mW. This corresponds to $S_{\epsilon\epsilon}  = 0.8 \times 10^{-17}\,\mathrm{Hz}^{-1}$, i.e., more than three orders of magnitude lower than in the present experiment. We notice however that the intensity noise directly measured on our cooling beam is already lower than the $S_{\epsilon\epsilon} $ deduced from the behavior of the (0,2) mode peak area. The intracavity excess intensity fluctuations may be caused by pointing noise, a technical issue that can be solved with a careful analysis of the optical path.

The excess frequency noise is a bigger problem. At present, we show that it limits the cooling performance for the (0,1) mode, but it is likely to similarly affect the (0,2) mode as soon the amplitude noise is reduced. For this second mode, a frequency noise below $0.7 \times 10^{-2}\, \mathrm{Hz^2/Hz}$ is indeed required to drop $\neff$ below 100. A possible strategy to mitigate the effect of the excess frequency noise is to reduce its length and consequently increase the optomechanical coupling rate, at the price of giving up the design of a cavity without piezoelectric transducer. By decreasing the input mirror transmission down to 100 ppm we could shorten the cavity by a factor of 20 maintaining an over-coupled cavity with the mechanical modes in the resolved sidebands regime (thus assuring $\nBA < 1$). A phonon occupancy of around 7 seems then achievable with current oscillator parameters and laser excess noise, at least for the (0,2) mode. The effective quality factor would however drop to $\sim 3$, an unreasonable value if we consider the presence of the other mechanical modes and the low-frequency background. A goal of $\neff$ below 20, dominated by the thermal noise, is realistically within reach. 
To reach the so-called ground state level $\neff \simeq 1$, the quality factor of the membrane modes should then be increased above $10^8$, a threshold already achieved in thinner membranes isolated from the frame by tethers \cite{Norte2016} or patterned phononic crystal \cite{Saarinen2023,Tsaturyan2017}.   

The effect of excess frequency noise is equivalent to that of fluctuations in the cavity length, with spectral density $\,S_{LL} = (\Lb/\nu_{\mathrm{L}})^2 S_{\nu\nu}\,$. We have ascribed the observed effect to laser excess noise, but fluctuations in the length of the cavity spacer and in the position of the surface of the input coupler could play the same role. In this case, the discussed scaling with the cavity length is not valid, and we remark that the length fluctuations of a shorter cavity with piezoelectric transducer cannot be easily predicted. We note that a frequency noise spectrum of $10^{-2}\,\mathrm{Hz}^2/\mathrm{Hz}$ is equivalent to length fluctuations with a spectral density of $6 \times 10^{-34}\,\mathrm{m}^2/\mathrm{Hz}$ for the cavity used in this work. To reduce $\neff$ below 20, the cavity length noise should stay below $10^{-35}\,\mathrm{m}^2/\mathrm{Hz}$, and to approach $\neff = 1$, besides the mentioned higher $Q$, we require a further order of magnitude reduction in $S_{LL}$. 
In \cite{Pontin2018} we report an upper limit to the displacement noise of $\sim 10^{-36}\,\mathrm{m}^2/\mathrm{Hz}$ around 170 kHz for a 1.5 mm long cavity, including a piezoelectric transducer, at cryogenic temperature. At room temperature, the mechanical modes of the bulky input mirror give a structured spectral background varying between $10^{-36}$ and $10^{-34}\,\mathrm{m}^2/\mathrm{Hz}$ in our frequency region of interest \cite{Arcizet2008}. The calculated broadband Brownian noise of the mirror is around $10^{-37}\,\mathrm{m}^2/\mathrm{Hz}$ \cite{Levin1998}, and similar or lower levels are expected for thermoelastic and thermorefractive noise \cite{Braginsky1999,Braginsky2000,Gorodetsky2008}.
The target is therefore not out of reach, but it requires a careful evaluation of the displacement noise background.

\section*{Acknowledgements}
Research performed within the Project QuaSeRT funded by the QuantERA ERA-NET Cofund in Quantum Technologies implemented within the European Union's Horizon 2020 Program. We also acknowledge financial support from: PNRR MUR
Project No. PE0000023-NQSTI.

\bibliography{database}

\end{document}